\documentclass[12pt]{elsart}
 
\usepackage{latexsym}
\usepackage{amsmath} 
\usepackage{amsfonts} 
\usepackage{epsfig} 
 
\newcommand{\tdiff}[2]{\frac{d #1}{d #2}}

\newcommand{\expval}[1]{\langle #1 \rangle}

\newcommand{\abs}[1]{\left| #1 \right|}

\newcommand{\bea}{\begin{eqnarray}}
\newcommand{\eea}{\end{eqnarray}}
\newcommand{\beann}{\begin{eqnarray*}}
\newcommand{\eeann}{\end{eqnarray*}}
\newcommand{\trace}[1]{{\rm Tr} \left\{ #1 \right\}}
\newcommand{\order}[1]{{\mathcal O}\left( #1 \right)}

\newcommand{\timo}{\hat{\mathcal T}}

\newcommand{\nn}{\nonumber}

\begin{document} 

\begin{frontmatter}
\title{Dynamical Casimir Effect in a Designed Leaky Cavity}
\author[TU]{Gernot Schaller\thanksref{mail},}
\author[TU,UBC]{Ralf Sch\"utzhold,}
\author[TU]{G\"unter Plunien,} 
\author[TU]{and Gerhard Soff}
\address[TU]{Institut f\"ur Theoretische Physik, Technische  Universit\"at
Dresden, D-01062  Dresden, Germany}
\address[UBC]{Department of Physics and Astronomy,
University of British Columbia, Vancouver B.C., V6T 1Z1 Canada}
\thanks[mail]{Corresponding author, electronic address : 
{\tt schaller@theory.phy.tu-dresden.de}}
\date{\today}

\begin{center}
\end{center}

\begin{abstract} 
  The phenomenon of particle creation within a resonantly
  vibrating lossy cavity is investigated for the example
  of a massless scalar field at finite temperature. Leakage is provided
  by insertion of a dispersive mirror into a larger ideal cavity.
  Via the rotating wave approximation we demonstrate that for the 
  case of parametric resonance the exponential growth of the number 
  of created particles and the strong enhancement at finite
  temperatures are preserved in the presence of reasonable losses. 
  The relevance for experimental tests of quantum
  radiation via the dynamical Casimir effect is addressed.
\end{abstract}    
\end{frontmatter}

\begin{figure}[ht]
  \centerline{\mbox{\epsfxsize=7cm\epsffile{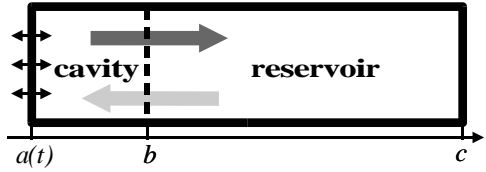}}} 
  \caption{\label{Fsystem}
  Model of a leaky cavity. A large
  ideal cavity is split up by a dispersive mirror into a lossy cavity
  and a reservoir. The left (ideal) wall of the cavity is vibrating.} 
\end{figure}

One of the most impressive manifestations of the nontrivial vacuum
structure in quantum field theory is the Casimir effect. 
More than fifty years ago Casimir \cite{casimir} predicted 
that two conducting parallel plates placed in the
vacuum will experience an attractive force. 
This effect has been verified experimentally, see e.g.\ 
\cite{lamoreauxmohideenbressi}. However, its dynamic counterpart with one 
or both mirrors moving and thus inducing phenomena like the effect 
of particle creation out of the vacuum has not yet been observed 
experimentally. This striking effect has been discussed by many 
authors (for a review see e.g.\ 
\cite{dodonov,tremblingcav,finitetemp,dalvit} and references
therein). It has been shown that under resonance conditions 
(i.e., when one of the boundaries undergoes harmonic oscillations 
at twice the frequency of one of the eigenmodes of the cavity) the 
phenomenon of parametric resonance will occur. In the case of an ideal 
cavity (i.e., one with perfectly reflecting mirrors) this leads to an 
exponential growth of the particle occupation numbers of the resonance 
modes \cite{dodonov,tremblingcav,finitetemp,dalvit}.
In this case a resonantly oscillating boundary is known to lead to squeezing 
of the vacuum state causing the creation of particles inside.  
\\
In view of this result an experimental verification of the dynamical
Casimir effect in principle appears to be rather simple 
-- provided the cavity is vibrating at
the resonance frequency for a sufficiently long time. 
Of course, this point of view is too naive
since ideal cavities do not exist. Therefore
it is essential to include effects of losses. Corresponding 
investigations have been performed for example in \cite{jrrad} 
based on the conformal invariance of the scalar
field in 1+1 space-time dimensions, see \cite{davies}. 
However, these considerations are a priori restricted to 1+1 dimensions and
cannot simply be generalized to higher dimensions. 
In 3+1 dimensions the character of quantum radiation 
(in particular the resonance conditions, see e.g.\ \cite{dodonov})
differs drastically from the 1+1 dimensional situation since the
spectrum of the eigenfrequencies is not equidistant anymore.
More realistic cavities with losses were considered in  
\cite{leakydodonov}. There effects of losses were
taken into account by virtue of a (static) master equation ansatz and
were not derived starting from first principles.
\\
In addition it is necessary to examine the effects of a detuned
external vibration frequency \cite{dalvit,leakydodonov}, since in an
experiment this frequency will always deviate from a desired value
\cite{forthcoming}.
\\
Furthermore most papers did not include temperature effects.  
The canonical approach has proven to be quite successful and
straightforward and it is also capable of including temperature
effects. It has been demonstrated in \cite{finitetemp} that these
corrections even enhance the effect of particle production in 
the case of an ideal cavity. However, the canonical approach still 
lacks a generalization for leaky cavities.
\\
The aim of the current article \cite{forthcoming} is to examine the
generic properties of the dynamical Casimir effect in a 
non-ideal, resonantly vibrating cavity.
For that purpose we consider a scalar field inside a dynamical leaky
cavity, which serves as a suitable model system. 
One simple way of constructing a leaky system is to insert a 
dispersive mirror into an ideal 3 dimensional cavity thus forming 
two leaky cavities coupled to each other. 
The left cavity is then bounded by a perfect mirror at $x=a(t)$ and a
dispersive mirror at $x=b$. It will be considered the leaky cavity,
whereas the right one (in addition bounded by a perfect mirror at $x=c$) is
understood as the (larger) reservoir, see also Fig.~\ref{Fsystem}.
Similar -- but static -- Fabry-Perot systems have been considered
frequently in quantum optics, see e.g.\ \cite{scully}. 
\\
Note that we are assuming a finite reservoir with a discrete spectrum 
instead of an infinite one leading to a continuum of modes. 
Since in any experimental setup the vibrating cavity will be surrounded 
by walls, etc., this assumption should be justified -- 
cf.\ the remarks after Eq.~(\ref{Eeffhv}) -- 
or at least the experiment can be designed in this way. 
\\
We consider a massless neutral scalar field coupled to the 
external potential [$\hbar=c=k_{\rm B}=1$ throughout]
\bea\label{Elagrangian}
  {\mathcal L} = \frac{1}{2} 
	\left[\partial_\mu \Phi(\mbox{\boldmath $r$};t)\right]
	\left[\partial^\mu \Phi(\mbox{\boldmath $r$};t)\right]
        - V(\mbox{\boldmath $r$};t) \Phi^2(\mbox{\boldmath $r$};t)\,, 
\eea
where the potential $V(\mbox{\boldmath $r$};t)$ simulates the ideal 
-- inducing Dirichlet boundary conditions -- and the dispersive mirrors. 
To keep the calculations simple we use an idealized $\delta$-type potential 
as considered in \cite{calogeracos} for the internal dispersive mirror
\bea 
  V(x;t) = \left\{ 
         \begin{array}{ll} \gamma \delta(x-b)&\mbox{if $a(t)<x<c$}\\ 
                           \infty & \mbox{otherwise} 
        \end{array} 
      \right. \, .
\eea
Note that when considering photons a dispersive mirror can be realized
by a thin dielectric slab with a very large
dielectric constant. This slab could be approximated by a space-dependent
permittivity $\varepsilon (x) = 1 + \alpha \delta (x-b)$ leading to
a similar eigenmode equation, see e.g.\ \cite{scully}. However, the
generalization to real photon creation in dynamic Fabry-Perot
cavities is not entirely straightforward (cf.\ \cite{finitetemp}) 
and remains subject to further elaborations.
\\
According to \cite{calogeracos} the parameter $\gamma$ enters the 
transmission and reflection amplitudes at frequency $\omega$ via 
${\mathcal T} = {\omega}/(\omega+i \gamma)$ and
${\mathcal R} = {-i\gamma}/(\omega+i\gamma)$. 
We expand the field operator into a complete and orthonormal set of
eigenfunctions 
$\hat\Phi (\mbox{\boldmath $r$},t) = \sum_\mu \hat{Q}_\mu (t)
f_\mu (\mbox{\boldmath $r$};t)$ satisfying the differential equation
$\{2 V(\mbox{\boldmath $r$};t) -\Delta\}f_\mu(\mbox{\boldmath $r$};t) 
= \Omega_\mu^2(t) f_\mu(\mbox{\boldmath $r$};t)$. 
A separation of variables leads to an equation accounting for the 
$x$-dependence of the eigenfunctions, which can be solved with the
ansatz (see also\cite{scully})
\bea\label{Eansatz}
  f_\mu^x (x) = \left\{ 
       \begin{array}{ll} 
        L_\mu \sin[\Omega_\mu^x (x-a)]\;:& \mbox{$a(t)<x<b$ (cavity)}\\ 
        R_\mu \sin[\Omega_\mu^x (c-x)]\;:& \mbox{$b<x<c$ (reservoir)}
       \end{array} 
       \right.\,.\!\!\!\!\!\!\!\!\!\!
\nn\\
\eea
Together with the matching conditions at $x=b$
this leads to a transcendental equation for the eigenfrequencies
\bea\label{Eequation}
-\frac{2\gamma}{\Omega_\mu^x} = \cot\left[\Omega_\mu^x(b-a)\right] + 
        \cot\left[\Omega_\mu^x(c-b)\right]\,.
\eea
For now we will assume the internal mirror to have a large reflectivity, 
with the parameters $\eta_\mu=\Omega_\mu^x/\gamma$ being small. 
Expanding in powers of $\eta_\mu$ one finds two sets of frequencies 
(with a relative shift of $\order{\eta_\mu}$ compared to those of ideal 
cavities), 
whose insertion in (\ref{Eansatz}) leads to two separate
sets of modes -- mainly concentrated in the system and the reservoir, 
respectively. We will denote the lowest left-dominated 
frequency obtained from (\ref{Eequation}) by $\Omega_L^x$ and to avoid
confusion we will also introduce the fundamental perturbation
parameter $\eta=\eta_L=\Omega_L^x/\gamma$.
In the following \cite{forthcoming} we shall refer to the different modes as 
left-dominated modes [large in the leaky cavity and small in the
reservoir: $R_\mu^{\rm (left)} = \order{\eta}$] and right-dominated modes 
[large in the reservoir and small in the leaky cavity: 
$L_\mu^{\rm (right)} = \order{\eta}$].
Note that when considering mirrors with a very high transparency one can
always increase the accuracy by including higher orders in $\eta$ or
even solving the associated equation (\ref{Eequation}) numerically.
\\
The total Hamiltonian for the scalar field is obtained from
(\ref{Elagrangian}) using a Legendre transformation. Via insertion of
the field operator eigenmode expansion $\hat{\Phi}$ it can be 
written as a sum (see e.g.\ \cite{tremblingcav}) of a free Hamiltonian 
\bea
  \hat{H}_0 = \frac{1}{2} \sum_\mu \hat{P}_\mu^2 
  + \frac{1}{2} \sum_\mu (\Omega_\mu^0)^2 \hat{Q}_\mu^2\,,
\eea
and a perturbation Hamiltonian $\hat{H}_I = \hat{H}_I^S + \hat{H}_I^V$ with
\bea
  \hat{H}_I^S = \frac{1}{2} \sum_\mu \Delta \Omega_\mu^2 (t) 
  \hat{Q}_\mu^2\,,\quad
  \hat{H}_I^V  =  \sum_{\mu\nu} 
  \hat{P}_\mu M_{\mu\nu} (t) \hat{Q}_\nu\,.
\eea
Adopting the nomenclature of \cite{tremblingcav} we refer to the two
parts of the perturbation Hamiltonian as the squeezing term
$\hat{H}_I^S$ and the velocity contribution $\hat{H}_I^V$.
The quantity $\Delta \Omega_\mu^2(t) = \Omega_\mu^2(t)-(\Omega_\mu^0)^2$ 
denotes the deviation of the time-dependent eigenfrequencies 
$\Omega_\mu^2(t)$ from the unperturbed ones $(\Omega_\mu^0)^2$.
The inter-mode coupling matrix 
$M_{\mu\nu}(t) = \int d^3 r \dot{f}_\mu (\mbox{\boldmath $r$};t)
f_\nu (\mbox{\boldmath $r$};t)$ is anti-symmetric
due to the ortho-normality of the modes and the
Dirichlet boundary conditions at the ideal mirrors, see
e.g.\ \cite{tremblingcav}. Note that both factors 
vanish for a stationary cavity. For the case of parametric resonance 
this complicated Hamiltonian can be simplified considerably by
applying the rotating wave approximation (RWA), see e.g.\ 
\cite{finitetemp,law}.
Let us assume the resonance case where during the time interval 
$[0,T]$ the left boundary of the cavity undergoes harmonic oscillations 
\bea
a(t) = a_0 + \epsilon (b-a_0) \sin(\omega t)\,,
\eea
with a small amplitude $0<\epsilon \ll 1$ and the external vibration
frequency $\omega$. With the duration of the vibration being
sufficiently long $\omega T \gg 1$ -- i.e., after many
oscillations -- one can approximate the time evolution operator in the 
interaction picture  
\bea\label{Etime}
  \hat{U}(T,0) = \timo \exp\left[-i\int_{0}^{T} \left(
                \hat{H}_I^S (t)+\hat{H}_I^V (t) \right) dt\right]
\eea
in the following way:
\\
Each term in the series expansion of (\ref{Etime}) can be rewritten 
as a multiple product of Hamiltonians without time-ordering
and terms involving commutators with Heaviside step functions. The
latter terms yield strongly oscillating integrands
\cite{finitetemp,forthcoming} and their contribution to (\ref{Etime})
is therefore comparably small. Performing the time-averaging they will
be neglected within the RWA as if one
would naively neglect the time ordering completely. 
\\
The remaining multi-integrals factorize and can be expanded into
powers of $\epsilon \ll 1$ and $(\omega T) \gg 1$. Within the RWA all
terms of $\order{\epsilon^J(\omega T)^K}$ with $J>K$ are
neglected. (Note that terms with $J<K$ do not occur.) 
The contributions with $J=K$ -- i.e., exactly the
terms in which the oscillations of the creation and annihilation
operators $\sim \exp(i\Omega_{\rm op} t)$ [interaction picture]
are compensated by the external time dependence 
[$\Delta\Omega_\mu^2(t)\sim\sin(\omega t)$ and 
$M_{\mu\nu}(t)\sim\cos(\omega t)$] -- will be
kept. In the resonance case ($\omega=\Omega_{\rm op}$) the relevant time
integrations can be approximated by  
\bea\label{Eint}
  \int_0^T \epsilon \omega e^{i \omega t} 
        \left\{\begin{array}{c}
        \sin(\omega t) \\
        \cos(\omega t) 
        \end{array}\right\} dt
  \stackrel{\rm RWA}{=}\frac{1}{2}\epsilon\omega T 
        \left\{\begin{array}{c}
        i \\
        1
        \end{array}\right\}\,.
\eea
Accordingly, the remaining terms in (\ref{Etime}) can be re-summated to
yield an effective time-evolution operator 
$\hat{U}_{\rm eff}(T,0)=\exp(-i\hat{H}_{\rm eff}^I T)$ with the
effective interaction Hamiltonian
$\hat{H}_{\rm eff}^I T \stackrel{\rm RWA}{=} \int_0^T \hat{H}_I^S(t) dt
+\int_0^T \hat{H}_I^V(t) dt$. 
\\
This effective interaction Hamiltonian can be calculated as follows:
In terms of annihilation and creation operators the squeezing interaction
Hamiltonian $\hat{H}_I^S$ decomposes into a sum of terms like 
$\Delta\Omega_\mu^2(t)\hat{a}_\mu^{(\dagger)}(t)\hat{a}_\mu^{(\dagger)}(t)$,
where $\Delta \Omega_\mu^2 (t) = 
2 \epsilon (\Omega_{\mu}^{x0})^2 \sin(\omega t) + \order{\epsilon^2}$ 
in the resonance case. Due to the trivial time dependence of the 
creation and annihilation operators in the interaction picture
$\hat{a}_\mu(t)=\hat{a}_\mu \exp(-i\Omega_\mu^0 t)$ 
the time integration of $\hat{H}_I^S$ involves
many oscillating terms, whose time average is rather small compared 
to that of constant contributions.
Only the terms where the oscillation
of $\Delta \Omega_\mu^2 (t)$ compensates the oscillation of the
operators, i.e., where the squeezing resonance condition (see also
\cite{finitetemp,dalvit,leakydodonov})
\bea
  \omega = 2 \Omega_\mu^0
\eea
holds, will be kept within the RWA. 
The spectrum of the cavity is assumed to be well-separated, i.e., the
relative distance of the different energy levels of interest is much
larger than $\order{\epsilon}$.
We choose the frequency $\omega$ to be exactly twice
the unperturbed frequency of the lowest left-dominated mode
$\Omega_L^0$ (fundamental resonance).
\\
In the series expansion of (\ref{Etime}) this leads with the aid of 
(\ref{Eint}) to a time-averaged effective squeezing Hamiltonian 
\bea
  \hat{H}_{\rm eff}^S = 
  i \xi \left[(\hat{a}_L^\dagger)^2 - (\hat{a}^{\phantom{\dagger}}_L)^2\right]\,,
\eea
with the squeezing parameter 
$\xi = \epsilon\Omega_{L}^0 ({\Omega_{L}^{x 0}}/{\Omega_{L}^0})^2/4$.
\\ 
The same procedure can be applied for the velocity interaction Hamiltonian:
Here the coupling matrix $M_{\mu\nu}(t)$ factorizes into a
time-independent geometrical factor 
$m_{\mu\nu}=\int d^3r (\partial f_\mu/\partial a) f_\nu$ 
and the velocity of the boundary 
$M_{\mu\nu}(t)=m_{\mu\nu} \dot{a}(t) + \order{\epsilon^2}$.
However, here the occurrence of 
inter-mode couplings results in an additional different resonance 
condition (see also \cite{finitetemp,dalvit})
\bea\label{Eresvelocity}
  \omega & = & \abs{\Omega_{\mu}^0 \pm \Omega_{\nu}^0} \,, 
\eea
where we are mainly interested in $\mu=L$ \cite{forthcoming}. 
In general the above resonance condition cannot be fulfilled by a 
left-dominated mode $\nu$ -- just as for a perfect cavity \cite{kubus}. 
(The situation is completely different in 1+1 dimensions with
$\hat{H}_I^V$ always contributing due to the equidistant spectrum, 
see e.g.\ \cite{dodonov,finitetemp}.)
However, in contrast to an ideal cavity it
can still be fulfilled by some right-dominated mode $\nu=R$. 
Accordingly, one similarly finds with (\ref{Eint}) a time-averaged 
effective velocity Hamiltonian
\bea\label{Eeffhv}
  \hat{H}_{\rm eff}^V & = & i \chi  
  \left(\hat{a}_{L}^\dagger \hat{a}_{R}^{\phantom{\dagger}} 
	- \hat{a}_{L}^{\phantom{\dagger}}\hat{a}_{R}^\dagger\right)\,,
\eea
with $\chi=\epsilon\Omega_L^0\left(\sqrt{\Omega_R^0/\Omega_L^0}
+\sqrt{\Omega_L^0/\Omega_R^0}\right)m_{LR}(b-a_0)/4$ being the velocity 
parameter of the system. It follows from the characteristics of left- and
right-dominated modes that the geometry factor becomes small 
$m_{LR}=\order{\eta}$ in the limit of a nearly perfectly reflecting
mirror which implies that $\chi/\xi =\order{\eta} \ll 1$.
The resonance condition (\ref{Eresvelocity}) can in general be 
fulfilled by many modes -- but the main contribution to the particle creation
is induced by the squeezing of the $L$-mode which couples only to the 
right-dominated $R$-mode in our considerations. 
\\
Assuming an arbitrarily large reservoir one would of course obtain many 
$R$-modes coupling to the fundamental $L$-mode of interest.
However, for a finite length of the right cavity $c-b$ and a sufficiently
long vibration time $T \gg c-b$ the resonance condition will single out 
a finite number of $R$-modes only. 
Since the effect of losses is additive \cite{forthcoming} to lowest order 
in $\eta$ -- i.e., for high-quality cavities -- it is sufficient to consider
one $R$-mode only.    
\\
Accordingly, the total effective interaction Hamiltonian reads
\bea \label{Eheff}
  \hat{H}_{\rm eff}^I = 
        i\xi\left[(\hat{a}_L^\dagger)^2-(\hat{a}^{\phantom{\dagger}}_L)^2\right] 
        +i\chi\left[\hat{a}_L^\dagger \hat{a}^{\phantom{\dagger}}_R
        -\hat{a}^{\phantom{\dagger}}_L\hat{a}_R^\dagger \right]\,. 
\eea 
We want to calculate the expectation value of particle number
operators that are explicitly time-independent in the interaction
picture: 
\bea
  \expval{N_\mu(T)} \stackrel{\rm RWA}{=}
        \trace{ \hat{U}_{\rm eff}^\dagger(T,0) \hat{N}_\mu 
        \hat{U}_{\rm eff}(T,0) \hat{\rho}_0}\,, 
\eea
where $\hat{\rho}_0=\exp(-\beta\hat H_0)/Z$ denotes the initial
statistical operator of the canonical ensemble. 
\\
Since the $\hat{U}_{\rm eff}$ is unitary one can introduce new
perturbation time ($T$) dependent ladder operators 
\bea\label{Eneu}
  \hat{a}_\mu(T) = e^{+i \hat{H}_{\rm eff}^I T} \hat{a}_\mu 
                        e^{-i \hat{H}_{\rm eff}^I T}\,,
\eea
and solve for their $T$-dependence.
This can be done by defining a 4-dimensional column vector 
(cf.\ \cite{forthcoming,dalvit})
\bea
  \underline{\hat{x}} (T) = \left( 
                \hat{a}^{\phantom{\dagger}}_L (T) \,,\,  
                \hat{a}_L^\dagger (T)\,,\, 
                \hat{a}^{\phantom{\dagger}}_R (T)\,,\,
                \hat{a}_R^\dagger (T) 
                \right)^{\rm T}\,.
\eea
Via differentiating (\ref{Eneu}) one can derive an equation for 
$\underline{\hat{x}}$
\bea
  \tdiff{\underline{\hat{x}}}{T} = 
  i \left[\hat{H}_{\rm eff} , \underline{\hat{x}}(T) \right]
                = \underline{A}\, \underline{\hat{x}}(T) \,.
\eea
Since we restrict ourselves to two coupling modes and a quadratic
Hamiltonian, $\underline{A}$ is a $4\times4$ number-valued matrix.  
More sophisticated couplings will simply increase the dimension of
$\underline{A}$. This differential equation is then solved by the
formal evolution matrix $\underline{U}(T)$ 
\bea 
  \underline{\hat{x}}(T)= \exp\left(\underline{A}\,T\right) 
                \underline{\hat{x}}(0) 
                = \underline{U}\,(T) \underline{\hat{x}}(0)\,,
\eea
i.e., initial and final ladder operators are related via a
Bogoljubov transformation.
In the present case the effective interaction Hamiltonian
(\ref{Eheff}) implies a very simple form of $\underline{A}$ 
\bea
  {\underline{A}}  = \left( \begin{array}{cccc} 
                        0 & 2\xi & \chi & 0\\ 
                        2\xi & 0 & 0 & \chi\\
                        -\chi & 0 & 0 & 0\\
                        0 & -\chi & 0 & 0 
                     \end{array} \right)\,,
\eea
with $\lambda_i = \pm \xi \pm \sqrt{\xi^2-\chi^2}$ 
being its eigenvalues. The time evolution matrix 
$\underline{U}(T)=\exp(\underline{A} T)$ is omitted here for brevity
but can certainly be calculated, e.g. using some computer algebra system.
Considering the time evolution of the new annihilation
and creation operators 
$\hat{\underline{x}} (T) = \underline{U}\,(T) \hat{\underline{x}} (0)$ 
one finds that the expectation values of particle number operators -- in
particular $\expval{\hat{N}_{L}}=\expval{\hat{x}_2\hat{x}_1}$ 
-- can be calculated via inserting
$\hat{\underline{x}} (T)$ leading to a bilinear form. 
The full response function turns out to be a combination of matrix 
elements of $\underline{U}(T)$ \cite{forthcoming}
\bea \label{Efulln}
  \expval{N_{L}(T)} & = & (U_{12}U_{21} + U_{14}U_{23})\nn\\ 
                 &   & +(U_{11}U_{22} + U_{12}U_{21}) \expval{N_{L}^0}
                        \nn\\ 
                 &   & +(U_{13}U_{24} + U_{14}U_{23})\expval{N_R^0}\,,
\eea
and similarly for particles created in the reservoir with 
$\expval{\hat{N}_R}=\expval{\hat{x}_4\hat{x}_3}$. 
The temperature dependence is taken into account by assuming a
Bose-Einstein distribution for the initial particle occupation numbers 
\bea
  \expval{N^0_\mu} = \expval{N^0_{R/L}} = \frac{1}{\exp(\beta\Omega^0_\mu)-1} 
\eea
of the resonance modes. 
Taking the limit of $\eta \to 0$ (ideal cavity) one
recovers the results found by other authors \cite{dodonov,finitetemp,dalvit}.
The explicit expression \cite{forthcoming} found for
$\expval{N_{L}(T)}$ does reflect a purely 
exponential growth of the particle numbers in the resonance modes 
as long as $\chi<\xi$, since the time-dependence of $\underline{U}(T)$ 
is governed by the eigenvalues of $\underline{A}$ via $\exp(\lambda_i T)$.
With $\chi=\order{\epsilon\Omega_L^0 \eta}$ this also leads to an
upper bound for the mirror transmittance $\eta$ above which
(corresponding to a highly transparent mirror) one
finds oscillations \cite{forthcoming}.
\\
In order to quantify the order of magnitudes let us specify the
relevant parameters:
A cavity with a typical size of $\Lambda \approx 1$ cm would have a
fundamental resonance frequency of $\Omega_{L}^0\approx 150$ GHz
i.e., the coupling right dominated mode has a
frequency of $\Omega_{R}^0=3\Omega_{L}^0\approx 450$ GHz. 
We will assume a dimensionless vibration amplitude of $\epsilon=10^{-8}$,
see also \cite{leakydodonov}.
\\
Consequently one would have to sustain the vibrations over an interval
of several milliseconds in order to create a significant number of 
particles. But even after only one millisecond 
-- i.e.\  $\approx 10^8$ periods -- a classical estimate based on a 
quality factor of $Q=10^8$ \cite{leakydodonov} would indicate drastic
energy losses. 
However, our calculations based on a 
complete quantum treatment show that the effects of losses 
do not drastically modify the exponential growth of the created
particles' number as long as $\eta \ll 1$.
Of course our calculations are based on the assumption that the larger 
cavity (including both reservoir and the leaky cavity) is
perfectly conducting and that the RWA-conditions such as  $T \gg c-b$ hold. 
The resulting error, however, is of
$\order{Q^{-2}}$ and therefore -- even classically -- certainly
negligible for $Q^2\gg10^8$, e.g.\ $Q=10^6$.
Consequently, the experimental verification of the dynamical Casimir
effect could be facilitated by a configuration where the vibrating
cavity is enclosed by a larger one.
A classical estimate of the quality factor $Q$ for our cavity yields  
$Q = 2 \pi / {{\mathcal{\abs{T}}}^2} 
= 2 \pi [1+(\gamma/\omega)^2] = \order{\eta^{-2}}$.
Accordingly, the assumption of $Q=10^6$, i.e.\ $\eta=\order{10^{-3}}$
is completely sufficient to justify our approximations. 
\\
Note that a cavity at finite temperatures is even 
advantageous provided the cavity is still nearly ideal for the 
characteristic thermal wavelength. 
\\
In summary the experimental verification of the dynamical Casimir
effect could become feasible with lower cavity quality factors than in
\cite{leakydodonov} (e.g.\ $Q=\order{10^6}$, which is experimentally
achievable) -- provided the shift of the resonance frequency by 
$-\eta/(2b-2a)$ with $\eta=\order{10^{-3}}$ is taken into account.

The authors are indebted to A.~Calogeracos, V.~V.~Dodonov, and 
D.~A.~R.~Dalvit for fruitful discussions. 
R.~S.\ is supported by the Alexander von Humboldt foundation and NSERC. 
Financial support by BMBF and GSI is gratefully acknowledged.


\begin{thebibliography}{99}
\bibitem{kubus}
In \cite{dalvit} it has been shown that for ideal cavities with
special cavity dimensions (a cubic one) a strong inter-mode
coupling can occur. In that case a much smaller growth rate of
the coupling modes was found, which is counterproductive in view of an
experiment and not considered here. 
Instead we propose a rectangular or cylindrical cavity with a 
transcendental ratio of the dimensions thus inhibiting couplings of
two left-dominated modes.
\end{thebibliography}
\end{document}